\documentclass[reprint,twocolumn,showpacs,preprintnumbers,superscriptaddress,amsmath,amssymb]{revtex4-2}
\usepackage{dcolumn}
\usepackage{bm}
\usepackage[english]{babel}
\usepackage[strings]{underscore}
\usepackage{color} 
\usepackage{hyperref} 
\usepackage{graphicx}
\usepackage{siunitx}
\usepackage{xcolor}
\usepackage{amsmath,amssymb}
\usepackage[T1]{fontenc}
\usepackage{lmodern}
\usepackage[utf8]{inputenc}
\usepackage[english]{babel}
\usepackage{natbib}
\usepackage{hyperref}
\usepackage{mathtools}
\usepackage{tipa}
\usepackage{float}

\hypersetup{colorlinks=true,citecolor={blue},linkcolor={blue},urlcolor={blue}}
\usepackage{ulem}

\begin{document}
\title{Moiré Photonic Crystals: from Fabric to Magic}%

\author{Marion Lavignac}
\affiliation{Ecole Centrale de Lyon, CNRS, INSA Lyon, \\ Universit\'e  Claude Bernard Lyon 1, CPE Lyon, CNRS, INL, UMR5270, Ecully 69130, France}

\author{Hai Son Nguyen}
\affiliation{CNRS-International-NTU-Thales Research Alliance (CINTRA), IRL 3288, Singapore 637553}
\affiliation{Ecole Centrale de Lyon, CNRS, INSA Lyon, \\ Universit\'e  Claude Bernard Lyon 1, CPE Lyon, CNRS, INL, UMR5270, Ecully 69130, France}

\author{Xavier Letartre}
\affiliation{Ecole Centrale de Lyon, CNRS, INSA Lyon, \\ Universit\'e  Claude Bernard Lyon 1, CPE Lyon, CNRS, INL, UMR5270, Ecully 69130, France}

\author{Ségolène Callard}
\affiliation{Ecole Centrale de Lyon, CNRS, INSA Lyon, \\ Universit\'e  Claude Bernard Lyon 1, CPE Lyon, CNRS, INL, UMR5270, Ecully 69130, France}

\author{Lydie Ferrier}
\email{lydie.ferrier@insa-lyon.fr}
\affiliation{Ecole Centrale de Lyon, CNRS, INSA Lyon, \\ Universit\'e  Claude Bernard Lyon 1, CPE Lyon, CNRS, INL, UMR5270, Ecully 69130, France}

\begin{abstract}
	Moiré patterns have recently become a very active field in nanophotonics. Those structures exhibit novel photonic properties unattainable with traditional photonic crystals. Especially, moiré magic configurations have been shown to allow intriguing slow light modes with zero group velocity. Starting from macroscopic moiré patterns in the everyday life, we will then shift to the subwavelength scale of moiré photonic crystals and detail some of their unusual properties.
\end{abstract}
    
\maketitle

\onecolumngrid

\begin{center}
\vspace{-1.0em}
\includegraphics[
  width=1\linewidth,
  height=15cm,
  keepaspectratio,
  clip
]{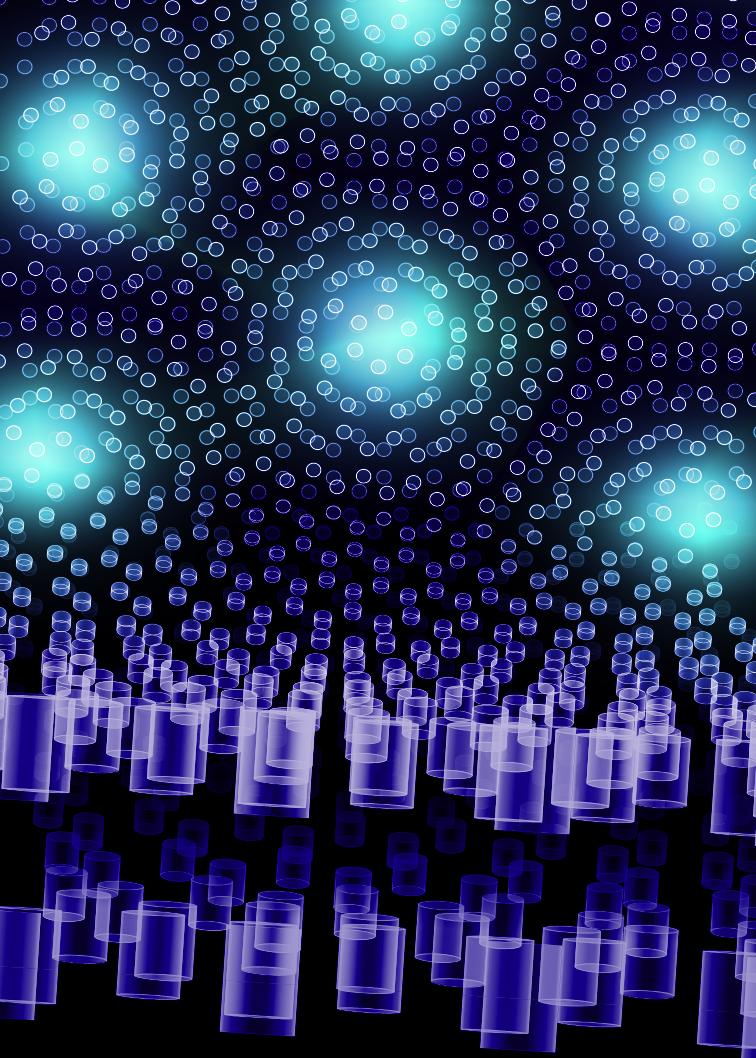}
\vspace{-0.6em}
\end{center}
\noindent\footnotesize
\noindent\rule{\linewidth}{0.4pt}
\textit{Editorial note:} A version of this manuscript was submitted to the “Perspective” section of \textit{Photoniques}, the French Optical Society’s bimonthly magazine for a broad readership. Owing to the magazine’s format, the published article is limited to seven references; this preprint provides a more complete bibliography.
\twocolumngrid
\begin{figure*}[!htbp]
    \centering
    \includegraphics[width=\linewidth]{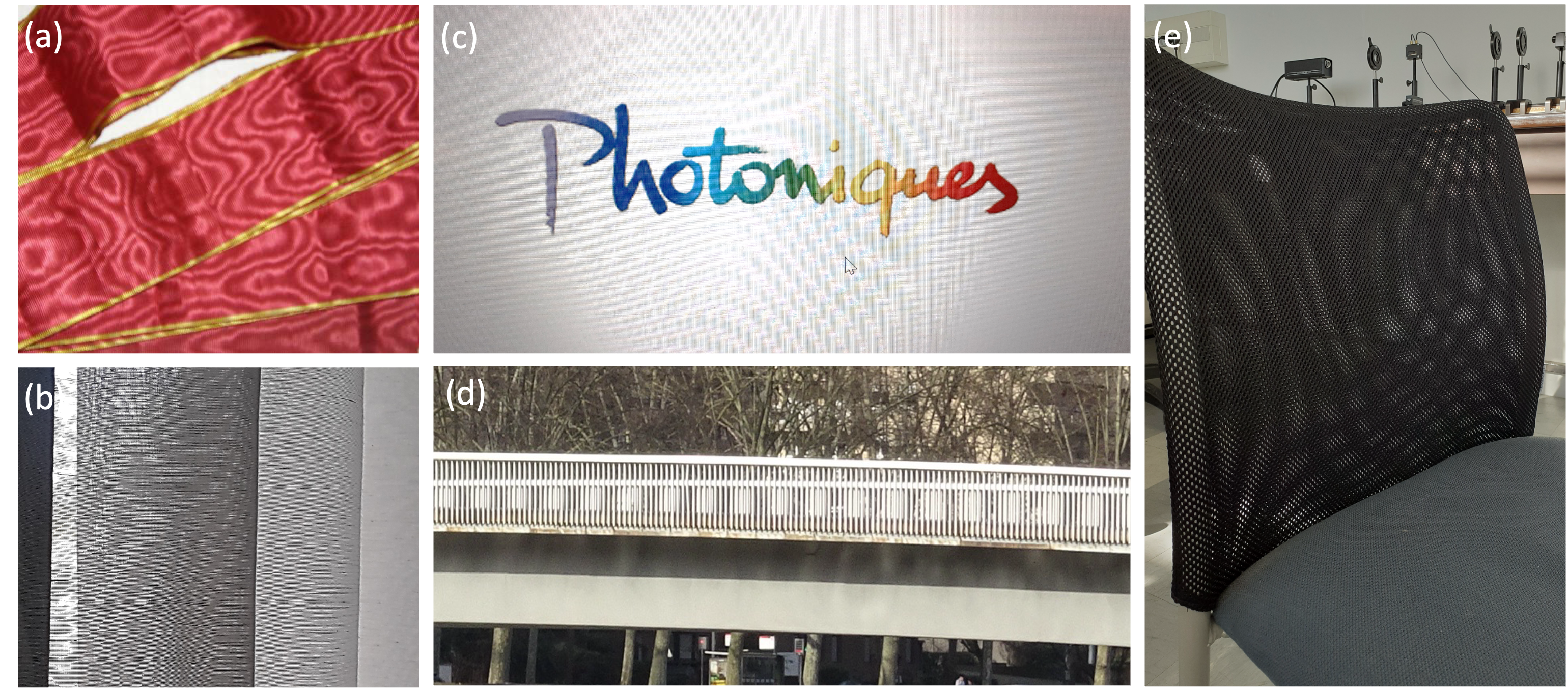}
    \caption{\textbf{Macroscopic moiré patterns.}
    (a) Moiré fabric.
    (b) Moiré patterns in curtains.
    (c) Moiré pattern from a resolution mismatch.
    (d) Striped moiré pattern produced by the superposition of bridge barriers (Pont Général Koenig, Lyon).
    (e) Moiré pattern on the back of a chair.
    \textit{Credits:} (a) User:Madame / Wikipedia Commons / \href{https://creativecommons.org/licenses/by-sa/4.0}{CC BY-SA 4.0}; (b-e) photos by Marion Lavignac.}
    \label{fig:moire-macroscopic}
\end{figure*}

\section{Introduction}
Have you ever noticed those weird wavy and flowing shapes magically appearing on your laptop screen when you try to take a picture of it with your smartphone? This is a moiré pattern. Originally, moiré \textipa{[mwa\textinvscr e]} was an adjective describing fabrics with wavy effects, obtained by strongly pressing the fabrics to flatten the fibres and modify light reflection. This technique was brought from Bagdad to England by the explorers from the 14th century, and reinvented during the 1740s by John Bagder who designed a machine named calender consisting of a heavy rectangular piece of rock used to press the fabrics. This invention gave the monopole of moiré fabrics in Europe to England. In 1753, the French government, worried by the French dependence on England’s moiré fabrics, invited John Bagder to settle in France, in Lyon, and transmit his knowledge to French weavers. Later, the calender underwent many developments and improvements, and Lyon became the French capital of moiré fabrics.
    
Moiré fabrics are not the only support to host moiré patterns. Similar wavy or beating shapes can be seen in curtains, on a bridge with grid barriers, or even if you take a picture of your computer’s screen with your smartphone (Figure~\ref{fig:moire-macroscopic}). In each case, the effect results from the superposition of two frames or lattices: two fabrics’ lattices, two grid barriers, the computer’s screen resolution and the picture resolution. Hence, moiré patterns extended to name every large-scale pattern formed by the superposition of two smaller-scale patterns.

\begin{figure*}[!htbp]
		\centering
		\includegraphics[width=1\linewidth]{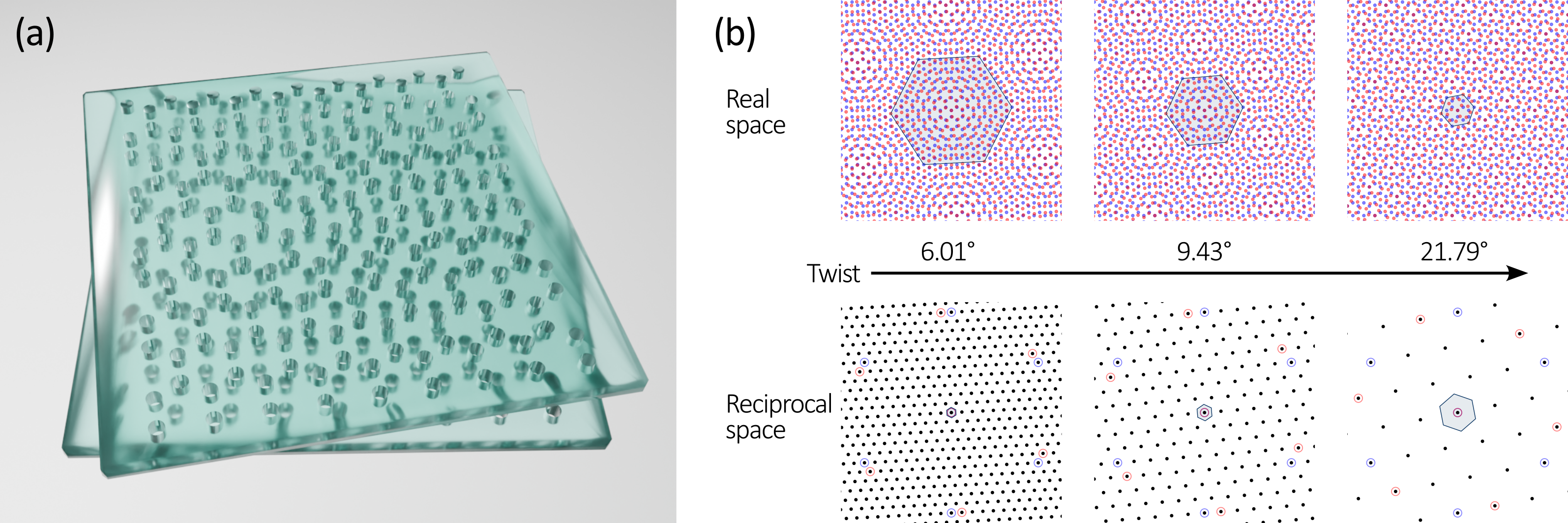}
		\caption{\textbf{Graphene-like moiré photonic crystal. }(a) Bilayer geometry. (b) Moiré lattice and reciprocal lattice depending on the twist angle.
		}
		\label{Fig.2}
\end{figure*}
    
 Today, moiré patterns are well-known in art and industry, sometimes desired and sometimes hated. Indeed, moiré effects can destroy the visual rendering of colour printing if the frames from the different colours have a particular alignment, or the display of pictures on LED screens due to the mismatch between the screen and the picture resolutions. On the other hand, the hypnotic aspect of their shapes has attracted artists’ attention, and took part in new artistic movements in the 1960s such as the so-called “optical art” or phase music. Moiré patterns also found applications in precise measurement: just as beating allows to tune two music instruments, moiré patterns can be used to align precisely two objects, or to measure their misalignment. Following this idea, they were applied to deformation and stress measurement, and to topography and 3D imaging with the technique called “projection fringes”. The level of precision that can be achieved is so high that similar methods are even used for precise alignment in nanofabrication processes such as some lithography techniques. Finally, moiré patterns were found to be useful for measuring thread density in fabrics using a type of striped ruler called lunometer, an invention that closed the loop with the textile industry.
 
 All the moiré patterns that have been mentioned so far are macroscopic. The underlying geometry is much larger than the wavelength of light in the visible range, and the ray optics framework applies. However, when the size is reduced down to the subwavelength scale, completely different and fascinating phenomena occur. This scale reduction was first introduced in a completely different field of physics. In 2018, it was discovered that, when two graphene sheets are stacked with a slight angular misalignment between their crystal lattices, bilayer graphene electronic properties could be tuned by varying the twist angle. More importantly, for some small and very precise angles ($\lesssim 1^\text{o}$) known as “magic angles”, twisted bilayer graphene becomes superconductive~\cite{Cao2018}. This discovery gave rise to twistronics, a new research field dedicated to twisted 2D materials. Since electrons in condensed matter crystals exhibit a wave-like character and behave similarly to photons in photonic crystals, parallels with photonics were quickly drawn, and research on moiré photonic crystals was launched.

 \section{Magic moiré photonic crystals}
 
 The exploration of moiré patterns in photonics started by emulating twisted bilayer graphene. Following this idea, each graphene sheet was modelled by a slab photonic crystal with a honeycomb lattice, consisting of a thin dielectric membrane that provides in-plane light confinement, and contains a subwavelength periodic modulation of the refractive index within the plane~\cite{Lou2021,Tang2021}. When the two layers are stacked with a small angular offset, a new large-scale hexagonal pattern appears: this corresponds to the moiré pattern~(Figure~\ref{Fig.2}.a).
 
 What happens geometrically can be viewed as a kind of spatial beating phenomenon. Beating in acoustics occurs when two sine waves of slightly different frequencies $\omega_0\pm\delta\omega$ are superposed. It results in an average sine wave at frequency $\omega_0$ slowly modulated at frequency $\delta\omega$. Similarly, in moiré photonic crystals, the temporal frequency offset $\delta\omega$ is replaced by a spatial offset arising from the twist angle between layers. This offset generates a long-range modulation that forms the moiré superlattice, whose period increases as the twist angle decreases. Just as with acoustic beats, the smaller the offset, the slower the modulation. This new periodicity is reflected in reciprocal space where a new Brillouin zone corresponding to the superlattice can be defined. It is smaller than the monolayers’ Brillouin zones, and shrinks as the twist angle is reduced (Figure~\ref{Fig.2}.b). However, it is important to notice that it is not always correct to talk about a “perfect” periodicity for the moiré pattern. In the general case, a moiré pattern is only quasi-periodic. This can be illustrated by the example of two square lattices twisted by 45°, for which no translational invariance exists due to $\sqrt2$ being irrational. Moiré patterns are strictly periodic only at specific discrete twist angles known as commensurate angles. At these angles, the perfect superposition of two lattice nodes at one point implies the existence of other perfectly aligned nodes elsewhere in the lattice. In such cases, the smallest distance between two points of perfect superposition defines the true lattice parameter of the superlattice. Nevertheless, it is always possible to refer to the moiré lattice corresponding to the pseudo-periodicity of the beating pattern.
 
\begin{figure}[!htbp]
		\centering
		\includegraphics[width=0.9\linewidth]{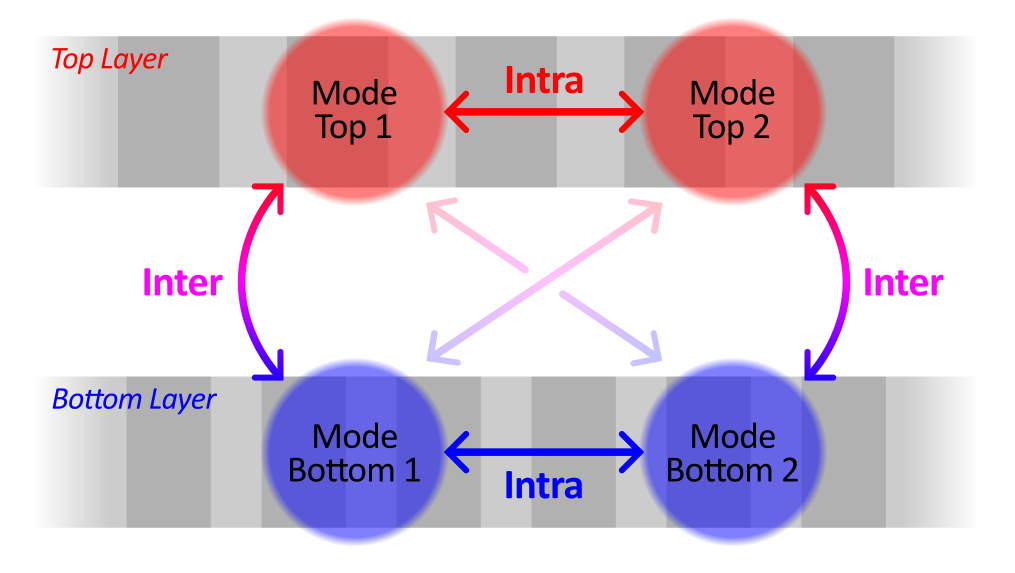}
		\caption{\textbf{Intra and interlayer couplings in a bilayer photonic crystal.} The modes can be guided modes living in each layer for example.}
		\label{Fig.3}
	\end{figure}

 How does light respond to these multiple periodicities? It depends on the distance separating the two layers. Qualitatively, if the interlayer spacing is larger than the operating wavelength, light will propagate through the two slab photonic crystals successively without noticing the moiré pattern. The observed effects are simply the combined influences of both layers individually, just as two successive gratings on an optical bench diffract light in two directions to form a 2D lattice of points in the far field. However, when the two layers are placed in near-field proximity, such that the interlayer distance is smaller than the wavelength, new behaviours emerge. In this configuration, light within the bilayer system not only perceives the periodicities of individual monolayers, but also experiences the moiré potential landscape, resulting in new photonic modes. In other words, the modes localized in each layer will interact with the (tilted) ones from the opposing layer. If the interlayer coupling between modes in different layers is of the same order of magnitude as the intralayer coupling within a single layer, the modes from each layer will strongly hybridize to form new bilayer modes (see Figure~\ref{Fig.3} for intra and interlayer couplings). These “moiré modes” inherit the characteristics of the moiré pattern, displaying a spatial profile that reflects the moiré (pseudo-)periodicity, and are highly sensitive to the twist angle.
    \begin{figure}[!htbp]
		\centering
		\includegraphics[width=0.8\linewidth]{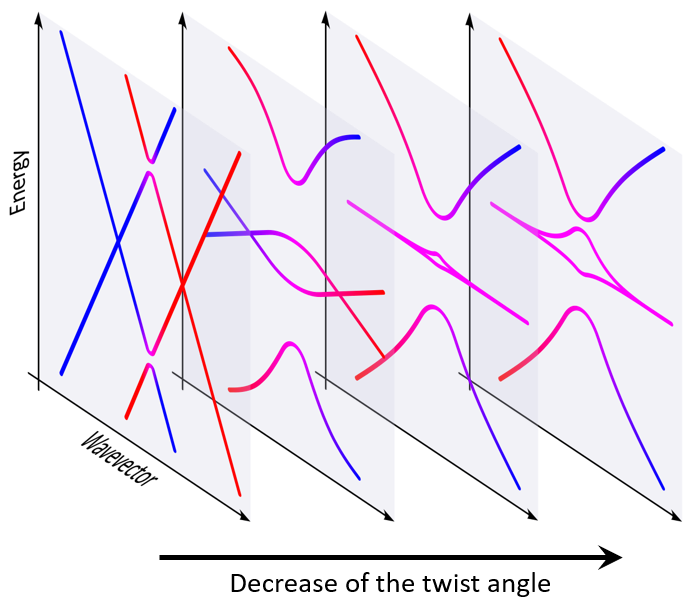}
		\caption{\textbf{Flat band formation within graphene-like geometry.}}
		\label{Fig.4}
	\end{figure}
 
 What has attracted attention on moiré photonic crystals is that, for a few very specific twist angles, some moiré modes see their field being strongly localized to small spots in the bilayer superlattice~\cite{Tang2021,Dong2021}. This behaviour is unusual because the geometry does not contain any heterostructure that could confine the modes so tightly. Instead, this effect is related to the moiré modulated interaction between the two layers. More specifically, it results from the interplay between the intra and interlayer coupling strengths, which determine the spatial profile of the bilayer moiré modes. The sharply localized modes are easy to identify in the band diagram of the bilayer photonic crystal: they are characterized by the emergence of a flat band, $\forall\vec{k}\mathrm{,\ \ }\omega(\vec{k})\mathrm{=}\omega_0$, that signifies zero group velocity across all wavevectors. This is why people refer to “slow light” in the context of moiré photonic systems. The twist angle plays a crucial role in the emergence of these slow modes. Changing the twist angle effectively shifts both lattices relative to each other, hence modifying the way two modes located in different layers interact together. Consequently, a slight change in the twist angle has a pronounced impact on the interlayer coupling, giving rise to “magic angles” that correspond to a particular balance between intra and interlayer coupling strengths. While looking at the formation process of the flat band as a function of the twist angle, it can be seen that it originates from the interaction between the slightly shifted dispersions of the two layers in k-space. Analogous to electrons in graphene, photonic dispersion in a honeycomb photonic crystal exhibits Dirac cones, a characteristic feature marked by a linear dispersion close to the K points, where the conduction and valence bands touch. At “magic angles”, the Dirac cones from the two layers strongly hybridize and merge into a quasi-flat band dispersion (Figure~\ref{Fig.4}).

 \section{New moiré geometries}
 
 The first studies on moiré photonic crystals with honeycomb lattices were very similar to those on 2D materials. While retaining the concept of the moiré patterns, photonic crystals physics allows much more freedom in the geometry and shape of the moiré patterns than condensed matter physics, which is constrained by the availability and stability of existing materials. Consequently, the bilayer photonic crystal platform opens the door to a wide variety of moiré patterns, as there is no reason to restrict our imagination to bilayer graphene-like geometries at small twist angles~\cite{Oudich2024}.
 
 The first deviation from bilayer graphene geometry lies in the order of magnitude of the twist angle. In photonics, magic angles are not always small, but can be as large as 22° for example~\cite{Yi2022}. This difference might be related to the interlayer coupling being stronger between slab photonic crystals than between 2D materials, because photonic modes have a longer evanescent tail than electronic orbitals. This coupling can be tuned by adjusting the interlayer distance: the closer the two layers, the stronger the interlayer coupling. However, contrary to what one might think, the strongest interlayer coupling is not necessarily optimal as it must be balanced with the intralayer coupling.
 
\begin{figure}[!htbp]
		\centering
		\includegraphics[width=1\linewidth]{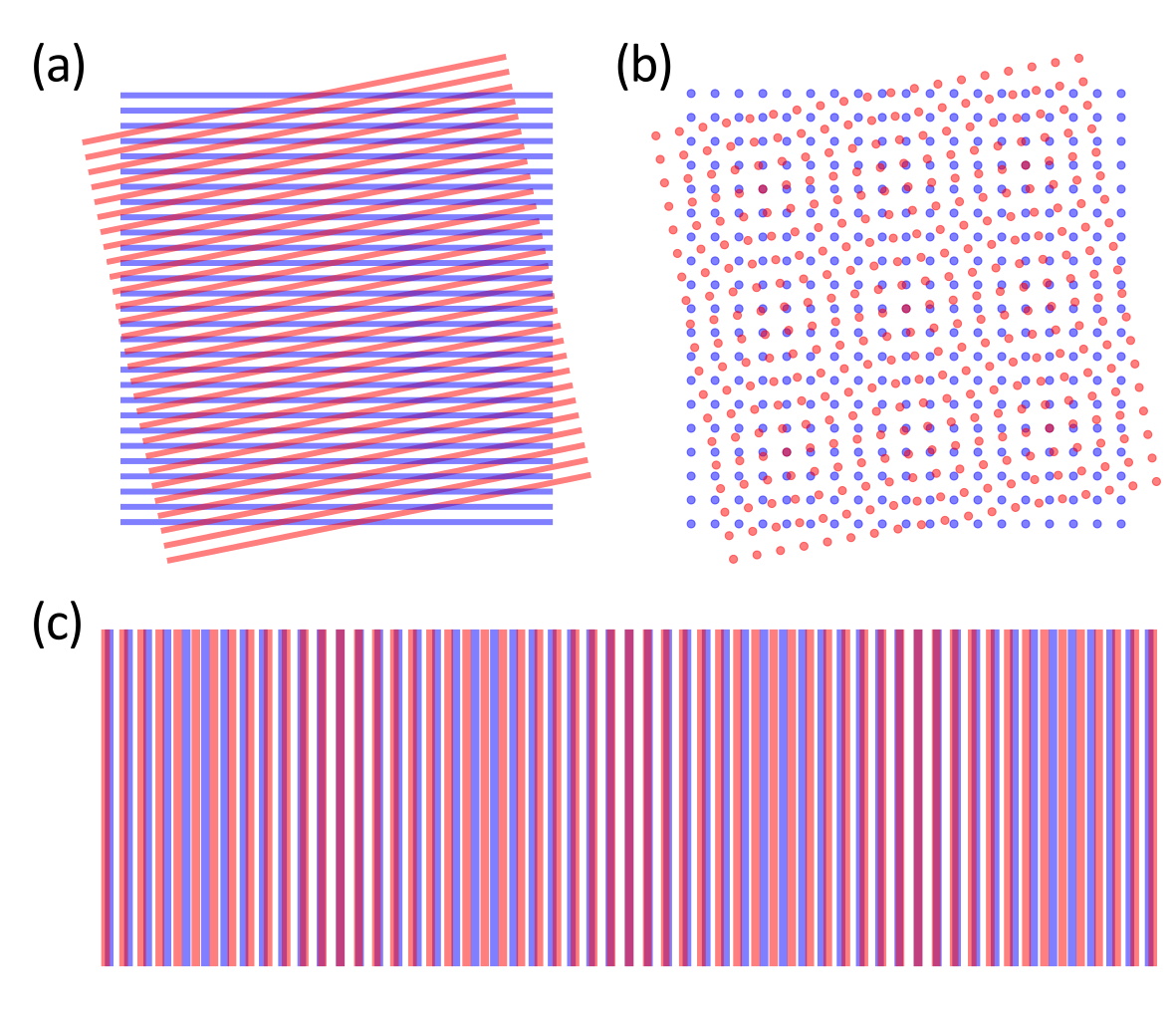}
		\caption{\textbf{Various moiré geometries.} (a) Twisted 1D gratings. (b) Twisted square lattices. (c) 1D gratings with lattice mismatch.}
		\label{Fig.5}
	\end{figure}

 Working with photonic crystals, completely different moiré geometries can also be explored. The lattices are not restricted to honeycomb structures, they can also be square, triangular, or even one-dimensional (1D). The mismatch parameter that generates the moiré pattern can arise either from a rotation or from a slight difference in lattice parameters. Figure~\ref{Fig.5} presents different types of moiré patterns. In particular, the moiré pattern made of two 1D photonic crystals with slightly different periods has been studied as a simplified system to explore the mechanism of flat bands formation~\cite{Nguyen2022,Saadi2024,Trushin2025}. It has been shown to exhibit flat bands as well as the graphene-like geometry for specific “magic configurations”. This 1D moiré pattern is strictly periodic with period $\mathrm{\Lambda}$ if the two periods $a_1$ and $a_2$ satisfy $\mathrm{\Lambda}\mathrm{=}(N+p)a_1\mathrm{=}Na_2$, where $N$ and $p$ are positive integers. For simplicity, $p$ is generally chosen to be equal to 1. The moiré number $N$ is the mismatch parameter that plays a role analogous to that of the twist angle. With this geometry, the balance between intra and interlayer couplings can be easily tuned, by adjusting the relative width of the rods and grooves and the interlayer distance respectively. This flexibility makes it easier to reach a “magic configuration”, regardless the value of the mismatch parameter $N$~\cite{Nguyen2022}.

\section{Properties}

 Initially, the main interest of photonic crystals was their capacity to control emission or absorption rates. An emitter emits photons at a rate proportional to the ability of the surrounding medium to support photonic modes, i.e. proportional to the photonic density of states. In free space, a photon with frequency $\omega_0$ must have a wavevector with $k\mathrm{=}\frac{\omega_0}{c}$, which limits the number of available states. In a moiré photonic crystal with a flat band at $\omega_0$, the same photon can be associated with a wide range of wavevectors, providing many more possible states. If the band is perfectly flat, the density of states even tends to infinity. Consequently, an emitter at $\omega_0$ embedded in this moiré photonic crystal can experience a strongly enhanced emission rate~\cite{Wang2025_QED}. This effect is opposite to the bandgap phenomenon that would prevent any emission of photon. This property makes moiré photonic crystals excellent candidates for compact devices requiring strong light-matter interactions such as laser sources.
 
 Even in the absence of a flat band mode, the tunability of moiré photonic crystals’ resonances and their sensitivity to slight geometrical changes make them very attractive. Moiré resonances are related to the moiré pattern, which can be strongly modified by adjusting the mismatch parameter and interlayer distance. These parameters represent additional degrees of freedom that, when carefully controlled, provide powerful means to tune the resonant wavelength or wavevector~\cite{Lou2022,Tang2023,Roy2025,Qu2025}, making moiré photonic crystals promising platforms for the design of tunable and reconfigurable optical devices.
 
 Finally, due to their bilayer structure and the symmetry breaking induced by the twist angle, most of moiré photonic crystals are chiral. Consequently, right- and left-circularly polarised light do not generally behave identically within this structure~\cite{Lou2021}. Therefore, moiré photonic crystals are able to support resonances that are selective to the polarisation handedness~\cite{Zhang2023,Zhang2025}. This property is highly relevant for applications such as detection and separation of enantiomers, or more generally for polarisation dependent light applications and enhanced chiral light-matter interactions. 
 
 To date, the vast majority of results on moiré photonic crystals come from theory and simulation. They are therefore largely restricted to commensurate configurations, since strictly periodic systems are easier to handle and to simulate numerically using standard simulation methods for photonic crystals. On the experimental side, achieving the structural precision required to observe flat bands and magic configurations remains challenging. The interlayer spacing and the mismatch parameter must be precisely controlled to observe a flat band. Then, to confirm that there is indeed a magic configuration, the spectral width of the band must be shown to pass through a minimum of almost zero when varying the mismatch parameter. One way to relax the fabrication constraints is to merge the two lattices into a single layer, i.e. to etch both photonic crystals in the same slab~\cite{Mao2021,Luan2023,Qin2024,Mou2025,Wang2025}. However, this approach loses control over the interlayer coupling and removes key bilayer properties. Finally, although quasi-flat bands have already been reported~\cite{Saadi2025,Spencer2025,Jing2025}, direct experimental demonstrations of true magic configurations are still lacking and their observation remains an active area of ongoing research.
 
 Despite these challenges, the field is rapidly evolving. Recent advances with moiré photonic crystals have demonstrated tunable optical sensors exploiting the additional degrees of freedom ~\cite{Tang2025}, as well as quantum well lasing ~\cite{Mao2021,Mou2025,Wang2025} or polariton lasing in perovskites ~\cite{Jin2025} benefiting from moiré flat bands, high density of states and sharp field localization. With the development of nanofabrication techniques and continued progress in experimental precision, direct observation of true magic configurations is becoming increasingly feasible. As experimental hurdles are overcome, moiré photonic crystals are poised to unlock new regimes of light-matter interaction, paving the way for novel quantum devices, enhanced optical sensors, and reconfigurable photonic platforms.
    
	\begin{acknowledgments}
     The authors acknowledge the French  National Research Agency (ANR), MILPHEUILLE (ANR-25-CE09-2554).
	\end{acknowledgments}

	\bibliography{Citations}

\end{document}